\documentclass[aps,prb,amsmath,twocolumn]{revtex4}

\usepackage{graphicx}
\usepackage{bm}

\begin{document}
\author{Artem V. Galaktionov}
\affiliation{I.E. Tamm Department of Theoretical Physics,
P.N. Lebedev Physical Institute, 119991 Moscow, Russia}
\affiliation{Laboratory of Cryogenic Nanoelectronics,
 Nizhny Novgorod State Technical University,
 603950 Nizhny Novgorod, Russia}
\author{Andrei D. Zaikin}
\affiliation{Institut f\"ur Nanotechnologie, Karlsruher Institut f\"ur 
Technologie (KIT), 76021 Karlsruhe, Germany} 
\affiliation{I.E. Tamm Department of
Theoretical Physics, P.N. Lebedev Physical Institute, 119991 Moscow,
Russia} 
\author{Leonid S. Kuzmin}
\affiliation{Chalmers University of Technology, Gothenburg, Sweden}
\affiliation{Laboratory of Cryogenic Nanoelectronics, Nizhny Novgorod
State Technical University, 603950 Nizhny Novgorod, Russia}

\title{Andreev interferometer with three superconducting electrodes}

\begin{abstract}
We develop a quasiclassical theory of Andreev interferometers with  three
superconducting electrodes. Provided tunneling interface resistance
between one superconducting electrode and the normal metal strongly
exceeds two others, significant current sensitivity to the external magnetic
flux is observed only at subgap voltages. If all barrier conductances are
comparable, multiple Andreev reflection comes into play and substantial
current modulation can be achieved in both subgap and overgap voltage
regimes. Our analysis reveals a large variety of interesting features which
can be used for performance optimization of Andreev interferometers.

\end{abstract}


\maketitle

\section{Introduction}
Andreev interferometers are often regarded as potential rivals to
Superconducting Quantum Interference Devices (SQUIDs) for a number of
applications with the possibility to achieve higher sensitivity and read-out
speed. Such kind of applications range from studying of switching
dynamics of individual magnetic nanoparticles to read-out of
superconducting qubits.

Almost   two decades ago Petrashov and co-workers performed
experimental analysis of magnetoresistance of mesoscopic
normal-superconducting (NS) hybrid structures \cite{Petr1}. This analysis
revealed conductance modulation greatly exceeding universal conductance
fluctuations. It was perceived, that this modulation should be attributed to
the effect of the phase difference $\chi$ between superconducting
elements of the hybrid structure influencing the process of Andreev
reflection (see Refs. \onlinecite{LR,bel} for a review). Since the normal bars
in the cross-like diffusive structures studied in experiments \cite{Petr1}
exceeded the superconducting coherence length $\xi_0$, the characteristic
energy scale for magnetoresistance modulation should be set by the
Thouless energy $\epsilon_{Th}=D/X^2$, where $D$ is the diffusion
coefficient and $X$ is the bar length in the cross.

In a diffusive NS structure with  good transmission of metallic interfaces the
magnitude of the magnetoresistance modulation is expected not to  exceed
few percents. This estimate is directly related to the so-called reentrance
effect in the corresponding NS structures reaching the maximum value
below 10 percent \cite{NazSt,GWZ}. If, however, tunnel barriers are present
in the NS system its conductance can change by much higher values
\cite{GWZ}. This is because Andreev conductance of a tunnel barrier at the
NS interface is much smaller than its normal state conductance while a
diffusive connector has the same low temperature Andreev conductance
as its normal one. Hence, one could expect that magnetoresistance
modulation could also be much more pronounced in hybrid NS structures
which contain tunnel barriers. This feature was indeed demonstrated
experimentally by Pothier {\it et al.} \cite{Poth} who observed the
maximum-to-minimum resistance ratio as a function of $\chi$ to be as high
as $\sim 5$. Physical insight into the phase dependence of the Andreev
conductance in hybrid NS structures is provided by generalization of the
Kirchhoff rules worked out by Nazarov \cite{Nazc}.

In the experiments discussed  so far at least one of the external electrodes
was in the normal state. Yet another option to fabricate the interference
device is to keep all available electrodes superconducting. In this way one
would be able to reduce dissipation.  Current harmonics of the current
would be generated in this case, which are multiples of the Josephson
frequency. However, they can be filtered out and the average current can
be measured which should reveal a dependence on the phase difference
$\chi$. Experiments with such structures were recently performed by
Meschke {et al.} \cite{MM}, and the corresponding theoretical analysis was
developed in Ref. \onlinecite{GT}.

\begin{figure}
\includegraphics[width=7cm]{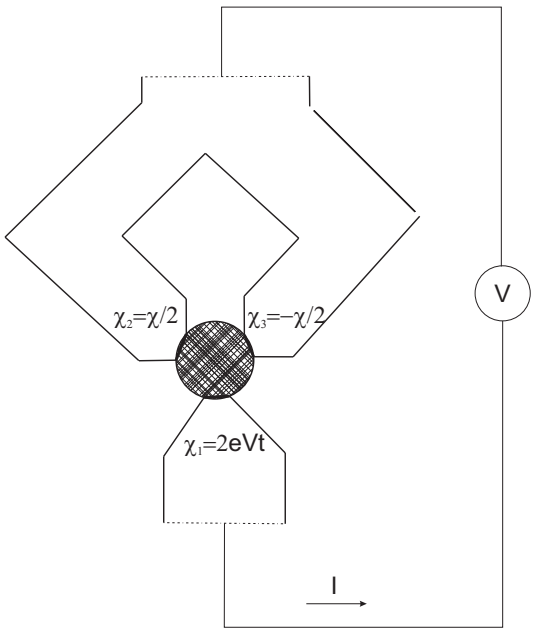}
\caption{Scheme of the setup: disordered normal insertion ("dot") between
three superconducting electrodes. The phase difference $\chi_2-\chi_3=\chi$ is
 caused by the magnetic flux piercing the loop.}
\label{setup}
\end{figure}

In this paper we study theoretically a different setup which is schematically
depicted in Fig. \ref{setup}. This device consists of a disordered normal
insertion embedded in-between three superconducting electrodes. Typical
size of this normal insertion $L$ does not exceed the superconducting
coherence length $\xi_0=\sqrt{D/\Delta}$, where $\Delta$ is the
superconducting gap of the electrodes, $D=v_Fl/3$ and $l$ is the elastic
electron mean free path. At the same time, the normal metal size obeys the
condition $L\gg l$. There exists a superconducting phase  difference $\chi$
between the electrodes 2 and 3 which is controlled by the external
magnetic flux $\Phi$ piercing the superconducting loop, i.e. $\chi =2\pi \Phi
/\Phi_0$, where $\Phi_0$ is the superconducting flux quantum.

\begin{figure}
\includegraphics[width=5cm]{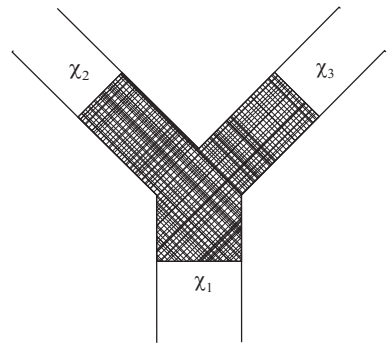}
\caption{Another possible realization of interferometer with the Y-shaped 
normal metal insertion.}\label{setup2}
\end{figure}

In what follows we will generally assume  that interfaces between normal
metallic dot and superconducting electrodes are weakly transmitting and
their normal state conductances $G_1$, $G_2$ and $G_3$ are supposed
to be smaller than the dot conductance $\sim \sigma_D {\cal A}/L$. Here
$\sigma_D=2e^2 DN_0$ is the Drude conductivity of the normal metal,
${\cal A}$ is the typical contact area between the normal metal and the
electrode and $N_0$ is the density of states at the Fermi-surface per spin
direction. This relationship between conductances assures the voltage
drops only across NS interfaces, while there exist no significant voltage
variations inside the normal dot. Under these conditions our results will not
depend on the particular shape of the normal metal insertion. For example,
one can equally apply our analysis to the setup displayed in Fig.
\ref{setup2}. We will also ignore charging effects which amounts to
assuming that all relevant charging energies remain much smaller than the
corresponding Josephson coupling energies \cite{SZ} and, in addition, that
tunneling conductances $G_{1,2,3}$ strongly exceed the quantum
conductance unit $G_q=e^2/h$ \cite{GZ10}. The latter condition also allows
to improve noise characteristics of the systems under consideration.

In the analysis \cite{MM,GT} it was assumed the contacts between
electrodes 2, 3 and the normal insertion are ideal i.e. highly transmitting)
while the first electrode is connected to the dot via a tunnel barrier.
Obviously in this case the conductances obey the condition $G_2\approx
G_3\gg G_1$. A proximity-induced $\chi$-dependent minigap inside the
normal dot develops in this case. As a result, the current-voltage
characteristics of such Andreev interferometer should be sensitive to the
phase difference $\chi$ which, in turn, can be controlled by external
magnetic flux $\Phi$. Qualitatively the same features hold in the situation
studied here as well. On the other hand, there also appear important
differences. For instance, we will demonstrate that the properties of the
Andreev interferometer essentially depend on the parameter $\gamma \sim
\tau_D\Delta$, where $\tau_D$ is the electron dwell time in the normal dot.
Similarly to the case of diffusive SNS junctions \cite{Bez} this parameter
effectively controls the strength of electron-hole dephasing in our device.
Another non-trivial feature of our structure is related to the effect of 
multiple Andreev reflections (MAR) \cite{mar} which becomes particularly 
important in the case $G_1\sim G_2\sim G_3$ providing significant modifications 
in the $\chi$-dependent current both in high and low voltage limits. The effect
of MAR on the properties of Y-shaped Andreev interferometers was studied
in Ref. \onlinecite{La} for a model system with normal insertion substituted
by a one channel quantum wire. Here we will address a realistic
configuration which includes many-channel diffusive conductors of an
arbitrary shape.

The structure of our paper is as follows. In section II we outline our general
formalism which allows to evaluate quasiclassical electron propagators for
the problem under consideration. Section III is devoted to the analysis of the
density  of states in the normal metallic dot as well as the flux-dependent
current across our structure in the regime $G_1\ll G_{2,3}$.
Non-perturbative MAR regime is studied in section IV. Our main
conclusions are summarized in section V.

\section{Quasiclassical analysis}
Our theoretical analysis is performed with the aid of Usadel equations (see,
e.g., Ref. \onlinecite{bel})
\begin{eqnarray}
&& -D\partial_ {\bm R}\left(\check g\circ  \partial_ {\bm R} \check g \right)+
\check\tau_z\frac{\partial \check g}{\partial t}+ \frac{\partial \check g}{
\partial t'}\check\tau_z +\label{useq}
\\ &&\left(-i\check \Delta(t) +ie\varphi(t)\right)\check g\ -\check g\left(
-i\check \Delta(t') +ie\varphi(t')\right)=0,\nonumber
\label{Usadel}
\end{eqnarray}
where $\check g$ represent the quasiclassical propagators that are
$4\times 4$ matrices depending on one spatial and two time variables
\begin{equation}
\check g ({\bm R},t,t')=\left( \begin{array}{cc} \hat g^R ({\bm R},t,t') &
\hat g^K ({\bm R},t,t')\\ 0 & \hat g^A ({\bm R},t,t')\end{array}\right).
\label{qp}
\end{equation}
Here $\hat g^{R,A,K}$ stand  for retarded, advanced and Keldysh
components respectively, each of them forming a $2\times 2$ matrix. The
electric potential is denoted by $\varphi({\bm R},t)$ and $e$ is the electron
charge. The product of propagators in Eq. (\ref{useq}) implies time
convolution
\begin{equation}
\left( \check g_1\circ\check g_2\right)(t,t')=\int\limits_{-\infty}^\infty
dt_1 \check g(t,t_1)\check g_2(t_1,t') d t_1.\nonumber
\end{equation}
The remaining matrices in Eq. (\ref{useq}) are defined as
\begin{equation}
\check \tau_z=\left(\begin{array}{cc} \hat\tau_z& 0
\\ 0& \hat\tau_z \end{array}\right),\;
\hat \tau_z=\left(\begin{array}{cc} 1& 0\\ 0& -1 \end{array}\right), \;
\check \Delta=
\left(\begin{array}{cc} \hat\Delta& 0\\ 0& \hat\Delta \end{array}\right)
\nonumber,
\end{equation}
where
\begin{equation}
\hat\Delta= \left(\begin{array}{cc} 0& \Delta({\bm R},t)
\\ -\Delta^*({\bm R},t)& 0 \end{array}\right) \label{sop}
\end{equation}
and $\Delta$ is the superconducting order parameter. Quasiclassical propagators
(\ref{qp}) also obey the normalization condition
\begin{equation}
\left( \check g\circ\check  g\right)(t_1,t_2)=\delta(t_1-t_2).\label{nc}
\end{equation}
In the immediate vicinity of a tunnel barrier Eqs. (\ref{useq}) are  not
applicable  and the quasiclassical propagators evaluated on both sides of
this barrier should be matched by appropriate boundary conditions \cite{KL}
\begin{equation}
2\sigma_D {\bm n} \check g \partial_{\bm R} \check g=\sigma_T
\left[ \check g_-,\check g_+\right].
\label{KLc}
 \end{equation}
Here the subscripts $\pm$ label the propagators on the right and the left
sides of the interface, ${\bm n}$ is the unit vector perpendicular to the
interface (directed from side "-" to side "+"). The combination in the
left-hand side of Eq. (\ref{KLc}) is continuous at the interface, the
commutator in the right-hand side is denoted by brackets and $\sigma_T$
stands for conductivity per unit square of the tunnel barrier.

It is convenient to Fourier transform the quasiclassical propagator with
respect to the time difference
\begin{equation}
\check g(\bm{R},\epsilon,t)=\int d t' \exp(i\epsilon t')
\check g\left(\bm{R},t+\frac{t'}{2}, t-\frac{t'}{2}\right).\label{ftp}
\end{equation}
Integrating Eq. (\ref{useq}) over  the normal metal volume ${\cal V}$ and
employing the Gauss theorem one finds
\begin{equation}
{\cal V}\left[ \epsilon\check \tau_z, \check{\overline{g}}\right]=
i\int d{\cal V} D\partial_k\left( \check g \partial_k \check g\right)=
iD \oint dS_k \left( \check g \partial_k \check g\right).\label{GO}
\end{equation}
With the aid of the boundary conditions (\ref{KLc})  from Eq. (\ref{GO}) we
obtain
\begin{equation}
[\check Z, \check{\overline{g}}]=0,\quad \check Z=
\sum_n \alpha_n \check g_n -i\gamma \frac{\epsilon}{|\Delta|}\check\tau_z,
\label{st1}
\end{equation}
where the unperturbed quasiclassical propagator of the $n$-the electrode
is  denoted by $\check g_n $ and we also introduced the notation
\begin{equation}
\alpha_n=\frac{G_n}{G_1+G_2+G_3}, \quad n=1,2,3.
\end{equation}
Finally, we introduce the parameter
\begin{equation}
\gamma=\frac{2\sigma_D {\cal V}|\Delta|}{D(G_1+G_2+G_3)} \sim
\tau_D |\Delta|\label{gdef}
\end{equation}
which is a direct generalization of the analogous parameter defined  for
diffusive SNS junctions \cite{Bez}. This parameter effectively controls the
strength of electron-hole dephasing in our system.

Combining Eq. (\ref{st1}) with the normalization  condition (\ref{nc}) we
arrive at the expression for the quasiclassical propagator in the normal dot.
It reads
\begin{equation}
\check{\overline{g}}=\frac{\check Z}{\sqrt{\check Z^2}}.\label{st2}
\end{equation}
We note that such an expression was previously  discussed in Ref.
\onlinecite{Str} in the context of the full-counting statistics and in Ref.
\onlinecite{Bez} in the context of MAR in SNS junctions.

\section{Density of states and tunneling current}

Let us first consider the limit $G_1\ll G_2, G_3$. In this case  Eq. (\ref{st1})
yields
\begin{equation}
\check Z=\frac{G_2\check g_2+G_3\check g_3}{G_2+G_3} -
i\gamma \frac{\epsilon}{|\Delta|}\check\tau_z.
\end{equation}
The retarded and advanced components of the equilibrium quasiclassical
propagators of the electrodes 2 and 3 are
\begin{equation}
\hat g^{R,A}(\epsilon)=\frac{\epsilon \hat\tau_z+\hat \Delta}{\xi^{R,A}},
\quad \xi^{R,A}=\pm\sqrt{(\epsilon\pm i\delta)^2-|\Delta|^2},\label{eqp}
\end{equation}
while the Keldysh component is defined  as $\hat g^K=\hat g^R F-F \hat
g^A$, where $F(\epsilon)=\tanh (\epsilon /2T)$ is the Fourier-transform of
the function $F(t)=-iT/\sinh [\pi T t]$. We also assume that there exists the
superconducting phase difference between the order parameters in the
electrodes 2 and 3, i.e. in these two electrodes we define
$\Delta=|\Delta|e^{\pm i\chi/2}$. As we already pointed out, this phase
difference is proportional to external magnetic flux $\Phi$ piercing the
superconducting ring.

Considering the diagonal component of the matrix  $(\hat g^R-\hat g^A)/2$
and employing Eq. (\ref{st2}) we recover the density  of states in the normal
dot equal to
\begin{equation}
n(\epsilon)={\rm Re} \frac{\epsilon}{\sqrt{\epsilon^2-\tilde \Delta^2}},
\label{d1}
\end{equation}
where we define
\begin{equation}
\tilde \Delta=\frac{\epsilon_g}{1+
\gamma\sqrt{1-\epsilon^2/|\Delta|^2}}.
\end{equation}
Here the parameter
\begin{equation}
\epsilon_g=|\Delta|\sqrt{1-\frac{4 G_2 G_3}{(G_2+G_3)^2}\sin^2\frac{\chi}{2}},
\label{bex}
\end{equation}
represents the minigap in the density of  states of the normal dot provided
the dephasing parameter tends to zero $\gamma \to 0$. In the symmetric
case $G_2=G_3$ Eqs. (\ref{d1})-(\ref{bex}) coincide with the
corresponding  expressions \cite{BSW}.

In order to interpret the result (\ref{bex}) in a  general case $G_2 \neq G_3$
we recollect that the maximum transmission value  $T_{\rm max}$ for the
system of two tunnel barriers with transmissions  $T_2$ and $T_3$ (both
much smaller than one) is defined by the well known formula
$$
T_{\rm max}=\frac{4 T_2 T_3}{(T_2+T_3)^2}.
$$
It follows immediately that Eq. (\ref{bex}) just defines the position of the
Andreev level with energy
$$\epsilon_A(\chi)=|\Delta|\sqrt{1-T_{\rm max}\sin^2(\chi/2)}.
$$

\begin{figure}
\includegraphics[width=7cm]{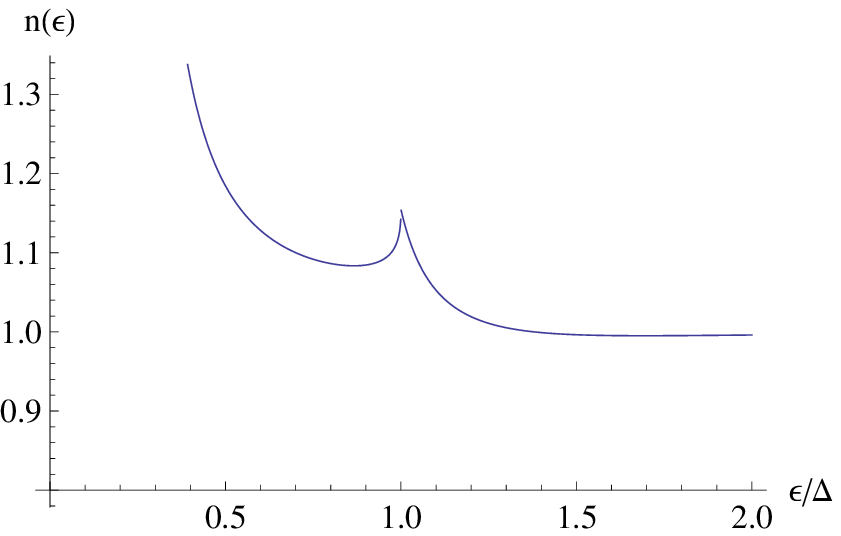}
\includegraphics[width=7cm]{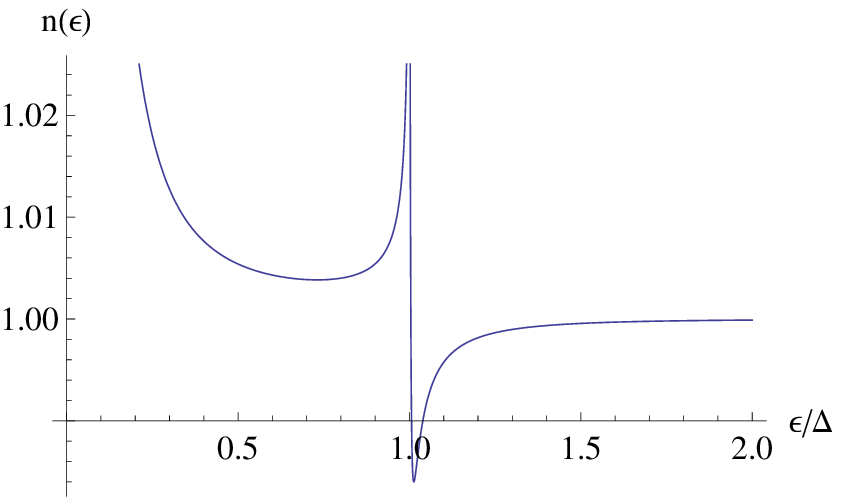}
\caption{(Color online) The density of states defined in Eqs.
 (\ref{d1})-(\ref{bex}) with $\epsilon_g =0.5|\Delta|$. The value of
 $\gamma$ is 1.0 for the upper plot and 10 for the lower plot.}
\label{densfig}
\end{figure}

For non-zero values $\gamma$ the minigap $\Delta_g$  in the normal dot
is obtained from the solution of the following equation
\begin{equation}
\Delta_g=\frac{\epsilon_g}{1+ \gamma\sqrt{1-\Delta_g^2/|\Delta|^2}},
\end{equation}
i.e. in the limit of strong dephasing $\gamma \gg 1$ the minigap  gets
reduced as $\Delta_g\approx \epsilon_g /\gamma$. The corresponding
density of states is exemplified in Fig. \ref{densfig}.

\begin{figure}
\includegraphics[width=8cm]{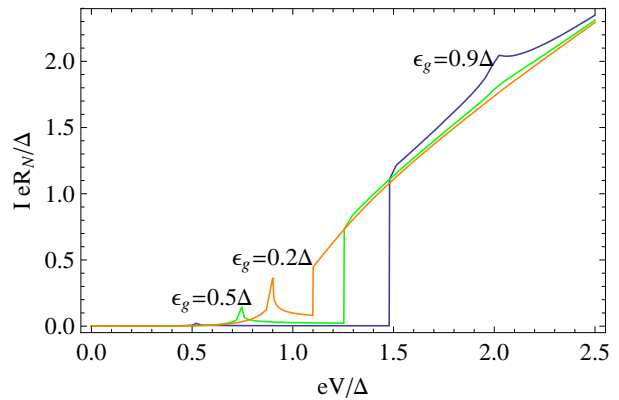}
\caption{(Color online) The current-voltage characteristics evaluated from
Eq. (\ref{tc}) for $T=0.1\Delta$ and $\gamma=1$. The values of $\epsilon_g$ are
0.9$\Delta$,  0.5$\Delta$ and 0.2$\Delta$. }
\label{tcvf}
\end{figure}

Let us  now evaluate the dissipative current flowing across our  device at a
given voltage bias $V$. In the limit $G_1 \ll G_{2,3}$ considered in this
section the voltage drop concentrates at the tunnel barrier between the first
electrode and the normal metallic dot. The current across this barrier reads
\begin{eqnarray}
&& I=\frac{1}{2e R_N}\int\limits_{-\infty}^{\infty} d\epsilon n(\epsilon)
n_0(\epsilon+eV)\times
\label{tc}\\ &&
\times\left(\tanh\frac{\epsilon+eV}{2T}-\tanh \frac{\epsilon}{2T}\right),
\nonumber
\end{eqnarray}
where $R_N \approx 1/G_1$  and
\begin{equation}
n_0(\epsilon)=\frac{|\epsilon|\theta\left(
 |\epsilon|-|\Delta|\right)}{\sqrt{\epsilon^2-|\Delta|^2}}
\end{equation}
is the standard BCS density of states in the  first  electrode, $\theta(x)$
denotes the Heaviside step function. Examples of the current-voltage
characteristics evaluated from Eq. (\ref{tc}) are displayed in Fig. \ref{tcvf}.
We observe that peculiarities (spikes) on the I-V curve occur at voltages
equal to $eV=\Delta-\Delta_g$, $eV=\Delta+\Delta_g$ and $eV=2\Delta$
(less pronounced). Here and afterwards $\Delta$ stands for the modulus of
the superconducting order parameter.

\begin{figure}
\includegraphics[width=8cm]{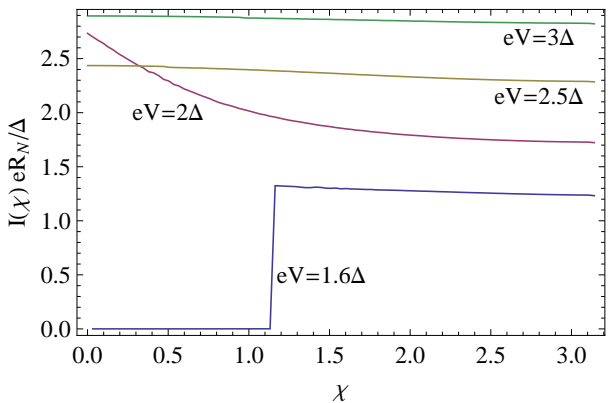}
\caption{(Color online) Phase-dependent tunneling current in a symmetric
device at $T=0.1\Delta$ and $\gamma=0.5$.}
\label{tmgr}
\end{figure}

Typical curves displaying the  phase dependence of the tunneling current
(\ref{tc}) at voltages $eV>\Delta$ are shown in Fig. \ref{tmgr}. We observe
that provided temperature is low and $eV<2\Delta$ there exists a
pronounced jump in the current which occurs as the phase $\chi$ reaches
the value corresponding to $eV=\Delta+\Delta_g(\chi)$. For $eV>2\Delta$
the phase dependence of the current is monotonous, the current
decreases with growing $\chi$  implying "positive magnetoresistance" of
our structure. The current modulation amplitude decreases rapidly as the
voltage increases and becomes almost negligible already at $eV \gtrsim
3\Delta$ .

\begin{figure}
\includegraphics[width=8cm]{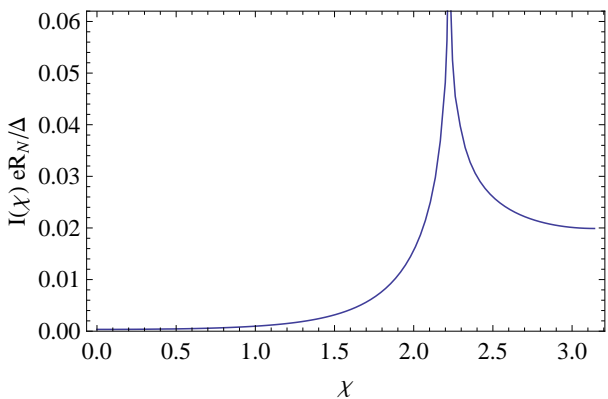}
\caption{(Color online) Phase-dependent tunneling current in a symmetric
device  at $T=0.1\Delta$, $\gamma=0.5$ and $eV=0.7\Delta$.}
\label{tmgrlt}
\end{figure}

As long as temperature is not  too low and the value $\exp (-\Delta/2T)$ is
not vanishingly small the $\chi$-dependent current is also observed at
voltages $eV<\Delta$. The  corresponding plot is shown in Fig. \ref{tmgrlt}.
One observes a pronounced current peak which occurs at
$eV=\Delta-\Delta_g(\chi)$.

In order to complete this part of our analysis we should add that  in order to
avoid hysteretic phenomena in our Andreev interferometer it is necessary
to obey the condition
\begin{equation}
\frac{{\cal L} I_C}{c}<\frac{\Phi_0}{2\pi}, \label{indl}
\end{equation}
which is exactly analogous to that well known for standard SQUIDs, see,
e.g., Ref. \onlinecite{Tinkh}. Here ${\cal L}$ is the loop inductance and
$I_C$ is the critical Josephson current in our device. The value $I_C$ can
roughly be estimated with the aid of the Ambegaokar-Baratoff formula for
the Josephson current
\begin{equation}
I_S(\chi)=\frac{\pi \Delta}{2e R_{23}} \sin\chi \tanh\frac{\Delta}{2T}
\label{AB}
\end{equation}
with $R_{23}=G_2^{-1}+G_3^{-1}$. Strongest deviations from this formula
occur for symmetric structures with $G_2=G_3$ provided $\gamma \to 0$.
In order to estimate these deviations in this case we can use the following
transmission distribution of conducting channels \cite{Bee}
\begin{equation}
P(T_n)\propto \frac{1}{T_n^{3/2}\sqrt{1-T_n}}.
\end{equation}
Employing this distribution and evaluating the Josephson  current, in the
limit  $T \to 0$ we obtain
\begin{equation}
I_S(\chi)=\frac{\Delta}{eR_{23}}\sin\chi\,{\rm K}\left( \sin^2\frac{\chi}{2}
\right). \label{nAB}
\end{equation}
Here ${\rm K} (x)=\int_0^{\pi/2} d\phi/\sqrt{1-x\sin^2\phi}$ is the  complete
elliptic integral. The maximum current value provided by Eq. (\ref{nAB}) is
higher than that determined from Eq. (\ref{AB}) by a factor $\approx 1.22$,
i.e. maximum deviations from Eq. (\ref{AB}) are in the range of 20 percents.

\section{Beyond tunneling limit: MAR regime}

In the previous section we restricted  the analysis of  I-V curves to the
tunneling limit, i.e. evaluated the current in the lowest order in the
conductance of the tunnel barrier with lowest transmission $G_1 \ll
G_{2,3}$. If this condition is violated, e.g., all three conductances have
approximately the same value $G_1 \approx G_2 \approx G_3$, it is
necessary to go beyond the lowest order perturbation theory and account
for higher order electron tunneling processes between superconducting
terminals. In this case an important role is played by the mechanism of
multiple Andreev reflection (MAR): quasiparticles with energies below the
superconducting gap $\Delta$ propagating inside the normal dot and
suffering Andreev reflections at different NS interfaces (i.e. being converted
from electrons to holes and back) are accelerated by the electric field and
eventually leave the dot area as soon as their energies exceed $\Delta$.
Recently it was demonstrated theoretically that this process essentially
influences the I-V curves of diffusive SNS junctions \cite{Bez}. The system
under consideration here is more complicated since it contains three
superconducting terminals.

Let us set the electrostatic potential of the first electrode equal to $-V$.
Then the quasiclassical electron propagator in this electrode reads
\begin{equation}
\check g_1(t,t')=e^{ieV\check \tau_z t}\check g_{{\rm eq}}(t-t')
e^{-ieV\check \tau_z t'},
\end{equation}
where the equilibrium  propagator $\check g_{{\rm eq}}(t)$ is defined in Eq.
(\ref{eqp}). The general expression for the current flowing into this electrode
through the tunneling barrier with conductance $G_1$ has the form
\begin{equation}
I(t)=\frac{\pi}{8e} G_1{\rm Tr}\left( \hat\tau_z \left[ \check{\overline{g}},
\check g_1\right]^{K}\right)(t,t),\label{currtr}
\end{equation}
where the  superscript $K$ denotes the Keldysh component of the
commutator. The electron propagator inside the normal dot
$\check{\overline{g}}$ is given by Eq. (\ref{st2}), where the matrix $\check
Z$ is defined in Eq. (\ref{st1}). One can also cast the expression for
$\check{\overline{g}}$ to the form
  \begin{equation}
\check{\overline{g}}=\frac{1}{\pi}\int\limits_{-\infty}^{\infty} d\lambda
\check K(\lambda), \quad \check K(\lambda)=\left( \check Z+
i\lambda\right)^{-1}.\label{stf}
\end{equation}
Provided all electrodes  are in the normal state the above equations just
yield Kirchhoff rules, i.e.
\begin{equation}
I=\frac{G_1}{G_1+G_2+G_3}\left( G_2(V_2-V_1)+G_3(V_3-V_1)\right).\label{kr}
\end{equation}
In what follows we will make use of this expression in order to normalize
the corresponding results derived below for the superconducting case.

It is convenient to rewrite  the  dependence of the quasiclassical
propagators (\ref{ftp}) on $\epsilon$ and $t$ with the aid of a series in
multiples of the Josephson frequency, i.e.
\begin{equation}
\check A(\epsilon,t)=\sum_{m=-\infty}^\infty \check A(\epsilon,m) e^{-2imeVt}.
\end{equation}
In this representation the matrix  $\check Z+i\lambda$ has only
$m=-1,0,1$ components which will be denoted as $\check
H_{-1}(\epsilon)$,  $\check H_0(\epsilon,\lambda)$ and $\check
H_{1}(\epsilon)$ respectively. The  condition
$$\left (\check Z+i\lambda\right)\circ\check K  =\delta(t-t')$$
reduces to
\begin{eqnarray}
&& \check H_0(\epsilon+meV,\lambda)\check K_\lambda(\epsilon,m)+
\label{invop}
\\ && \check H_1(\epsilon+(m-1)eV)\check K_\lambda(\epsilon-eV,m-1)+
\nonumber
\\&&  \check H_{-1}(\epsilon+(m+1)eV)\check K_\lambda(\epsilon+eV,m+1)=
\delta_{m,0}.\nonumber
\end{eqnarray}
Introducing $\tilde K_m(\epsilon,\lambda)=\check K_\lambda( \epsilon+meV, m)$,
one can rewrite Eq. (\ref{invop}) in the form
\begin{eqnarray}
&& \check H_0(\epsilon+2meV,\lambda)\tilde K_m(\epsilon,\lambda)+
\label{invstr}
\\&& \check H_1(\epsilon+(2m-1)eV)\tilde K_{m-1}(\epsilon,\lambda)+
\nonumber
\\&& \check H_{-1}(\epsilon+(2m+1)eV)\tilde K_{m+1}(\epsilon,\lambda)=
\delta_{m,0}.\nonumber
\end{eqnarray}
In order to resolve this equation one can employ the ansatz \cite{Bez}
\begin{eqnarray}
&& \tilde K_m=\check S_m \check S_{m-1} \ldots \check S_1 \tilde K_0;
\quad m>0,\\
&& \tilde K_m=\check P_m \check P_{m+1} \ldots \check P_{-1} \tilde K_0;
\quad m<0,\nonumber
\end{eqnarray}
with the aid of which Eq. (\ref{invstr}) yields recurrences relating 
$\check S_m$ and $\check S_{m+1}$
\begin{eqnarray}
&& \check S_m(\epsilon,\lambda)=-\left[ \check H_{-1}(\epsilon+(2m+1)eV)
\check S_{m+1}(\epsilon,\lambda) +\right.\nonumber \\ && \left.
\check H_0(\epsilon+2meV,\lambda)\right]^{-1} H_1(\epsilon+(2m-1)eV).
\label{rec1}
\end{eqnarray}
Similarly, for $m<0$, we obtain the following relationship between $\check
P_m$  and $\check P_{m-1}$
\begin{eqnarray}
&& \check P_m(\epsilon,\lambda)=-
\left[\check H_{1}(\epsilon+(2m-1)eV) \check P_{m-1}(\epsilon,\lambda) +\right.
\nonumber
\\ && \left.\check H_0(\epsilon+2meV,\lambda)\right]^{-1}
\check H_{-1}(\epsilon+(2m+1)eV).\label{rec2}
\end{eqnarray}
At $m=0$ we get from Eq. (\ref{invstr})
\begin{eqnarray}
&& \tilde K_0 (\epsilon,\lambda)=\left[  \check H_{1}(\epsilon-eV)
\check P_{-1}(\epsilon,\lambda)+\right.\\ && \left.\check H_{-1}(\epsilon+eV)
\check S_{1}(\epsilon,\lambda)+\check H_0(\epsilon,\lambda)\right]^{-1}.
\nonumber
\end{eqnarray}
Finally, we impose the "boundary conditions"
\begin{equation}
\lim_{m\rightarrow \infty} \check S_m=0,\quad
\lim_{m\rightarrow -\infty} \check P_m=0.
\end{equation}
Thus, proceeding  numerically we set $\check S_m,  \check P_m=0$  for
some large $|m|$ and then employ the recurrences (\ref{rec1}) and
(\ref{rec2}) in order to find $\check S_1, \check P_{-1}$ and  $\tilde K_0$.

Averaging of Eq. (\ref{currtr}) results in the following expression for the
current
\begin{eqnarray}
&&I=\int d\lambda\int d\epsilon {\rm Tr}
\bigg[ \check S_1(\epsilon,\lambda)\tilde K_0(\epsilon,\lambda)
\check L_{-1}(\epsilon+eV)+ \nonumber\\ && \check P_{-1}
(\epsilon,\lambda)\tilde K_0(\epsilon,\lambda)
\check L_{1}(\epsilon-eV) +\tilde K_0(\epsilon) \check L_0(\epsilon) \bigg],
\label{dint}
\end{eqnarray}
which demonstrates that our numerical procedure consists of  performing
the double integral in $\epsilon,\lambda$ and employing the matrix
recurrence relations (\ref{rec1}) and (\ref{rec2}) at each step of the
integration. The matrices depend on dimensionless parameters
$\gamma$,  $eV/\Delta$, $T/\Delta$, $\chi$ and $\alpha_{2,3}$ (the
parameter $\alpha_1$ is excluded by $\alpha_1=1-\alpha_2-\alpha_3$).

\begin{figure}
\includegraphics[width=8cm]{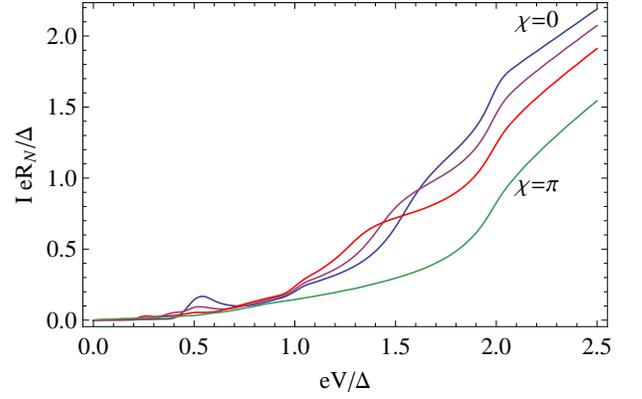}
\caption{(Color online) Current-voltage characteristics at
$\alpha_2=\alpha_3=0.45$, $T=0.1\Delta$ and  $\gamma=1.$ The values of
$\chi$ are 0, 1.3, 2 and $\pi$. The lower $\chi$ value corresponds to the upper
curve for $eV>2\Delta$. }
\label{tmar}
\end{figure}

\begin{figure}
\includegraphics[width=8cm]{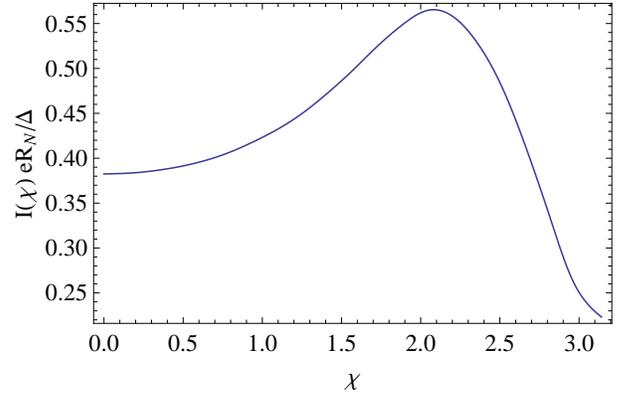}
\caption{(Color online) Phase-dependent current at
$\alpha_2=\alpha_3=0.45$, $T=0.1\Delta$,
$\gamma=1$  and $eV=1.3\Delta$. }
\label{ttmgr}
\end{figure}

Let us now  present some results of our numerical analysis of the problem
in question\cite{avail}. To begin with, it is satisfactory to observe that
several important features established in the lowest order in tunneling
survive -- though possibly with significant quantitative modifications -- also
within our non-perturbative analysis which includes the effects of MAR.
E.g., in Fig. \ref{tmar} we display the I-V curves evaluated at different values
of the phase $\chi$ for the value $G_1$ about 4.5  times smaller than
$G_2$ and $G_3$. All our results are normalized to the normal resistance
of the structure $R_N$ defined from Eq. (\ref{kr}) as
\begin{equation}
\frac{1}{R_N}=(\alpha_2+\alpha_3)(1-\alpha_2-\alpha_3)\left( G_1+G_2+G_3\right).
\end{equation}
Comparing the results presented in Fig. \ref{tmar} with those  obtained
perturbatively, cf.,  e.g., Fig. \ref{tcvf}, in both cases we observe peculiar
gap-like features at $eV=\Delta+\Delta_g$ in the I-V curves at voltages
$\Delta<eV<2\Delta $. These features are also qualitatively consistent with
the results \cite{La} obtained for a model of Y-shaped Andreev
interferometer with a normal part being substituted by a single mode
quantum wire. For instance, this simple model allows to predict current
peaks at voltages  $eV=\epsilon_A (\chi )$ and $eV=\Delta+\epsilon_A (\chi
)$, where  $eV=\epsilon_A (\chi )$ is the phase-dependent energy of the
Andreev subgap bound state. These peaks have the same physical origin
as those found here at $eV=\Delta_g (\chi )$ (cf. the curves displayed in
Fig. \ref{tmar} at $eV<\Delta$) and $eV=\Delta+\Delta_g (\chi )$, except in
our case the current peaks are modified both due to specific transmission
distribution in our structure and due to the influence of the parameter
$\gamma$.

A typical current-phase dependence is depicted in Fig. \ref{ttmgr}. This
curve can be qualitatively compared to one evaluated perturbatively and
presented in Fig. \ref{tmgr} at voltages $\Delta<eV<2\Delta$. Though these
curves differ quantitatively, their qualitative behavior remains somewhat
similar. Namely, in both cases we observe an increase of the current with
$\chi$ at smaller phase values followed by its decrease at larger values of
$\chi$.

\begin{figure}
\includegraphics[width=8cm]{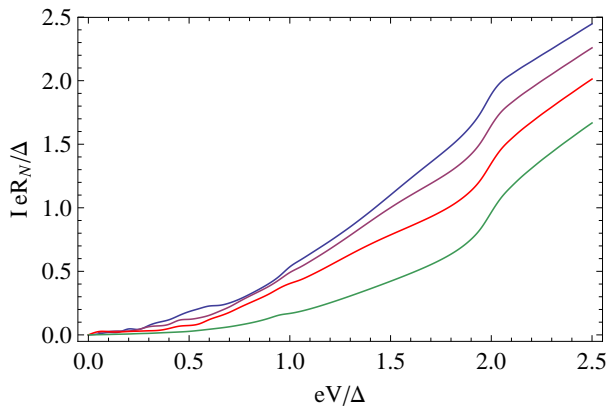}
\caption{(Color online) Current-voltage characteristics for
$\alpha_2=\alpha_3=1/3$, $\gamma=1$ and $T=0.1\Delta$ case. The values
of $\chi$ are 0, 1.3, 2 and $\pi$ (top to bottom).}
\label{vah}
\end{figure}

On the other hand, there also  exist  significant differences between the
results obtained in MAR and tunneling regimes. Perhaps the most essential
one is the presence of non-zero subgap current observed within the
non-perturbative regime even in the limit of low temperatures. Obviously
this feature is lacking within the lowest order perturbation theory in
tunneling. The subgap current is well pronounced at $G_1\approx G_2
\approx G_3$, as it is demonstrated in Fig. \ref{vah}. Similarly to the case of
SNS junctions \cite{Bez}, there exist somewhat chaotic phase -dependent
jumps of current in the subgap voltage region. At  $\chi=\pi$ the current is
strongly suppressed for subgap voltages $eV<\Delta$.

\begin{figure}
\includegraphics[width=8cm]{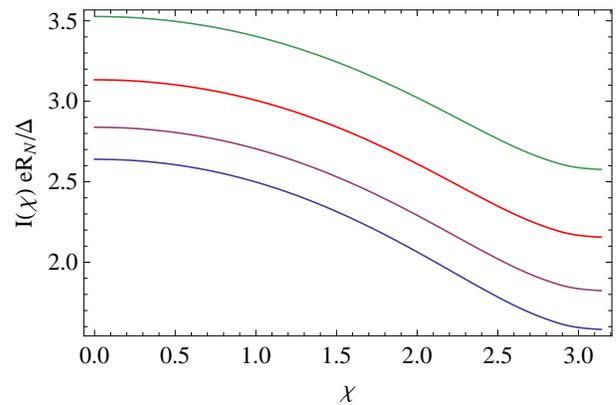}
\caption{(Color online) The phase-dependent current in the symmetric case
($\alpha_2=\alpha_2=1/3$) for $\gamma =0.5$, $T=0.1\Delta$
and $eV=2.3\Delta,\, 2.5\Delta, \, 2.8\Delta, \, 3.2\Delta$ (bottom to top).}
\label{hvph}
\end{figure}

Our results  demonstrate that MAR may essentially influence the phase
dependence of the current. In this respect it is instructive to compare Figs.
\ref{hvph} and \ref{tmgr}. While in the perturbative tunneling limit the current
modulation decreases rapidly with increasing voltage (see Fig. \ref{tmgr}),
there exist a clear voltage-independent modulation of the current in the
non-perturbative MAR regime (Fig. \ref{hvph}). This modulation is due to the
presence of a phase-dependent excess current with the amplitude $\sim
\Delta /(e R_N)$ which is not captured within the lowest order perturbation
theory in tunneling.

\begin{figure}
\includegraphics[width=8cm]{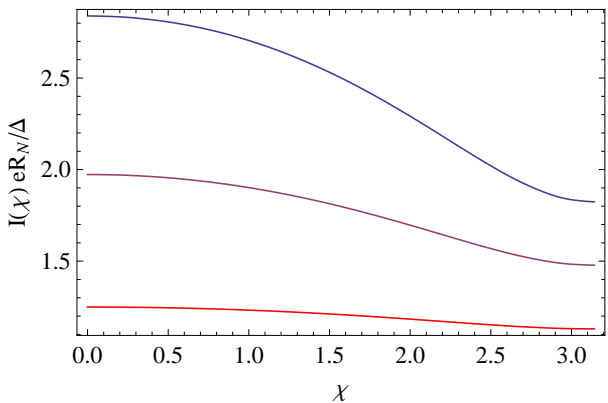}
\caption{(Color online) The phase-dependent current in the
symmetric case at $eV=2.5\Delta$, $T=0.1\Delta$  and
$\gamma=0.5, 2, 10$ (top to bottom).}
\label{gch}
\end{figure}

\begin{figure}
\includegraphics[width=8cm]{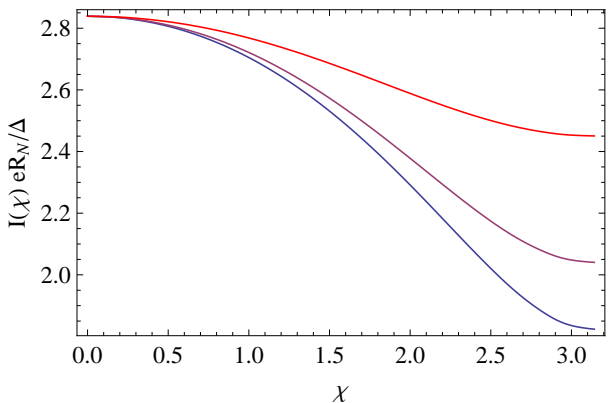}
\caption{(Color online) The phase-dependent current at $eV=2.5\Delta$,
$T=0.1\Delta$ and $\gamma=0.5$.  The lower, middle and upper curves correspond
respectively to $\alpha_2=\alpha_3=1/3$, to $\alpha_2=0.444$, $\alpha_3=0.222$,
and to $\alpha_2=0.555$, $\alpha_3=0.111$. }
\label{asch}
\end{figure}

Fig. \ref{gch} illustrates the dependence of the current modulation on the
electron-hole dephasing parameter $\gamma$. This modulation clearly
decreases with increasing $\gamma$ similarly to the minigap $\Delta_g$,
as it was discussed in the tunneling limit. It is worth pointing out that
conductance asymmetry $G_2\neq G_3$ also yields a decrease of the
current modulation, as it is demonstrated in Fig. \ref{asch}.

\begin{figure}
\includegraphics[width=8cm]{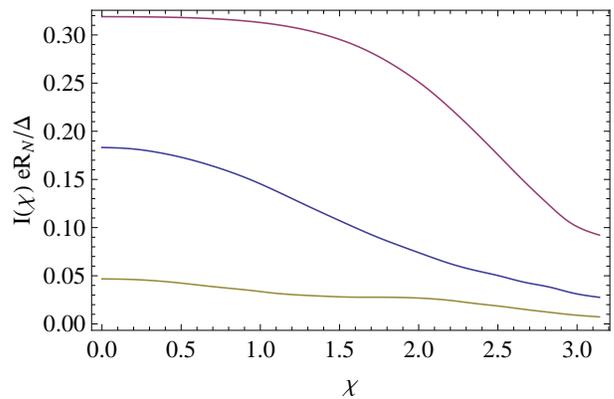}
\caption{(Color online) The phase-dependent current of the symmetric
junction at $T=0.1\Delta$ and $\gamma=1$. The voltage values are
$0.8\Delta$,  $0.5\Delta$ and $0.2\Delta$ (top to bottom).}
\label{mfig}
\end{figure}

The phase-dependent current at lower voltages is displayed in Fig.
\ref{mfig}. Comparing these results to those in Fig. \ref{tmgrlt} we observe
that the current takes much higher values in the non-perturbative MAR
regime. At the same time the current-phase dependencies turn out to be
considerably smoother in this regime. Note that by tuning the voltage value
one can reach the regime where the current depends monotonously on the
phase $\chi$, cf., e.g., the curve evaluated for $eV=0.5\Delta$. In this case
the current modulation by the factor $\sim 5$ is observed. This bias voltage
regime can be conveniently employed for magnetic flux measurements at
lower voltages. This regime appears advantageous as compared to, e.g.,
$eV=0.8 \Delta$ and $eV=0.2\Delta$, since in the latter cases there exist
extended flat regions with nearly $\chi$-independent current values.

\section{Conclusions}
In this paper we have constructed a  general theory of Andreev
interferometers with three superconducting electrodes. Our analysis
revealed a large variety of different regimes and features which can be
studied experimentally and used for performance optimization of these
devices.

In the case $G_1 \ll G_{2,3}$ the I-V curves can  be evaluated perturbatively
in the barrier transmissions. In this limiting case an important role is played
by the Andreev minigap $\Delta_g(\chi )$ which may cause strong
dependence of the current on the phase $\chi$ at not very large bias
voltages, as shown, e.g. in Figs. \ref{tmgr} and \ref{tmgrlt}. On the other
hand, current modulation decreases rapidly as $eV$ increases above
$2\Delta$, see Fig. \ref{tmgr}.

If all barrier conductances are comparable, $G_1 \approx G_2\approx
G_3$, it  is necessary to go beyond simple perturbation theory in tunneling
and include the effects of MAR into consideration. Features related to the
minigap $\Delta_g(\chi )$ persist also in this case, cf., e.g. Figs. \ref{tcvf}
and \ref{tmar}, but the curves become much smoother and the current
signal larger. Significant current-phase modulation can be achieved both in
subgap (Fig. \ref{mfig}) and overgap (Figs. \ref{hvph}-\ref{asch}) voltage
regimes, i.e. both these regimes can be used for successful operation of
Andreev interferometers with three superconducting electrodes. In the
regime of large voltages this modulation is due to phase-dependent excess
current which is not captured within the perturbative in tunneling analysis.

It is also important to add that both in perturbative and non-perturbative
regimes large values of the current modulation can be  achieved provided
the parameter $\gamma$ defined in Eq. (\ref{gdef}) remains sufficiently
small. This modulation decreases drastically for large values of $\gamma$
and, hence, such values should be avoided in Andreev interferometers.

Also we would like to make a remark concerning the effect  of current
noise. Although this effect deserves a separate analysis, it is clear already
at this stage that in the MAR regime it would be desirable to avoid working
in the limit of low voltages $eV\ll \Delta$, since in this case one could
expect dramatic increase of current noise which could compromise the
operation of Andreev interferometers. Theoretical analysis of current noise
correlator
\begin{equation}
{\cal S}(\omega)=\int d\tau e^{i\omega\tau} \left\langle\delta I(t) 
\delta I(t-\tau)\right\rangle, \quad \delta I(t)=I(t)-\langle I\rangle
\nonumber
\end{equation}
in the case of diffusive SNS junctions in the zero frequency  limit
\cite{Ave,Cuev} reveals that at low enough voltages one has ${\cal
S}/\langle I \rangle  \propto 1/V $, i.e. the signal-to-noise ratio should
decrease with decreasing voltage. This effect is directly related to MAR.
Indeed, an effective charge transferred at voltages  $eV\approx 2\Delta/n$
(with integer $n$) equals to $q=ne$. As the number of Andreev reflections
$n$ grows with decreasing voltage, the charge $q$ grows as $q\sim
(1+2\Delta/|eV|)$ and, hence, the ratio ${\cal S}/\langle I\rangle\propto q  $
grows too. We also remark that, as in the case of Andreev interferometers
with a normal electrode \cite{Reul}, one can also expect to observe noise
modulation depending on the phase difference $\chi$. However, at this
stage we do not expect that this effect could alter our conclusion about low
voltage regime being possibly problematic for successful operation of
Andreev interferometers with three superconducting electrodes.


\begin{thebibliography}{99}
\bibitem{avail} The C++ code of the numeric programs of this section is
    available upon request.
\end{thebibliography}
\end{document}